\documentclass[12pt,journal,comsoc]{IEEEtran}
\usepackage[T1]{fontenc}

\ifCLASSINFOpdf
\else
\fi
\usepackage{amsmath}
\usepackage[cmintegrals]{newtxmath}
\usepackage{graphicx}

\begin{document}
%
\title{Fusing Keys for Secret Communications: Towards Information-Theoretic Security}
%
%
%

\author{Longjiang~Li,~\IEEEmembership{Member,~IEEE,}
	~Bingchuan~Ma, 
	~Jianjun~Yang,
	~Yonggang~Li,
	and~Yuming~Mao%
	\thanks{L. Li is with the Department of Network Engineering,
		School of Information and Communication Engineering, University of Electronic Science and Technology of China, Chengdu, Sichuan, 611731 P.R. China e-mail: longjiangli@uestc.edu.cn}
	\thanks{B. Ma, J. Yang and Y. Mao are with University of Electronic Science and Technology of China, Chengdu 611731, China.}
	\thanks{Y. Li is with Chongqing University of Posts and Telecommunications, Chongqing 400065, China.}
	\thanks{Part of this work has been submitted to the IEEE for possible publication. Copyright may be transferred without notice, after which this version may no longer be accessible.}}

\maketitle

\begin{abstract}
Modern cryptography is essential to communication and information security for performing all kinds of security actions, such as encryption, authentication, and signature. However, the exposure possibility of keys poses a great threat to almost all modern cryptography. This article proposes a key-fusing framework, which enables a high resilience to key exposure by fusing multiple imperfect keys. The correctness of the scheme is strictly verified through a toy model that is general enough to abstract the physical-layer key generation (PLKG) mechanisms.  Analysis and results demonstrate that the proposed scheme can dramatically reduce secret outage probability, so that key sources with even high exposure probability can be practically beneficial for actual secret communication. Our framework paves the way for achieving information-theoretic security by integrating various key sources, such as physical layer key generation, lattice-based cryptography, and quantum cryptography.

\end{abstract}

\begin{IEEEkeywords}
Exposure-Resilient Cryptography; Physical-Layer Key Generation; Key-Fusing Transformations; Secret Outage Probability; Information-Theoretic Security; Physical Layer Security
\end{IEEEkeywords}

%
\IEEEpeerreviewmaketitle

\section{Introduction}
%
%
%
%

\IEEEPARstart{D}{ue} to the imminent threat of quantum computers, information-theoretic security \cite{Bloch2008} that emphasizes the strictest randomness typically measured in terms of min-entropy \cite{Dodis2011} has gained many attentions. Fruitful results, such as leakage-resilient cryptography \cite{Naor2009a} and quantum cryptography \cite{Arnon-Friedman2018}, have been achieved through modern cryptographic methods such as privacy amplification  and entropy accumulation. 
However, the exposure possibility of cryptographic keys still poses a great threat to almost all modern cryptography. Especially, 
the process of key agreement and distribution is often vulnerable to various attacks, such as side-channel attacks and cold boot attacks, which makes key exposure generally covert or unpredictable \cite{Canetti2000}\cite{Ni2019}. Typically, key exposure problem can be formulated as one-time attacks \cite{Naor2009a} or continuous leakage models \cite{Wu2019}, and the mainstream countermeasure is to periodically update potentially leaked keys by using fresh ones at high frequency \cite{Dodis2010, Zhou2019}, which results in wasting a large amount of precious entropy. 

In general, cryptographic keys can be generated through either secret key exchange or physical-layer wiretap codes \cite{Mathur2008, Liu2014}. Most of key exchange mechanisms, such as Diffie-Hellman key exchange algorithm (D-H), are based on some (unproved) difficulties of mathematical problems, e.g., factoring large integers or calculating discrete logarithms \cite{Maurer1993}. Physical-layer key generation (PLKG) is not dependent on those unproven hypothesis but relies on the condition that there are no eavesdroppers within half a wavelength radius of legitimate users \cite{Jiao2019}. Quantum cryptography or quantum key distribution (QKD) provides what is claimed to be the information-theoretic security based on quantum physics, but it faces a number of implementation challenges such as secret key rate, distance, cost and practical security \cite{Arnon-Friedman2018}. Lattice-based cryptography has the promise to resist quantum attacks, but it is still vulnerable to side channel attacks\cite{Chaudhary2019}. Therefore, it is highly desired to achieve information-theoretic security by integrating imperfect key sources that are easily obtained at affordable costs.

In this article, we propose a key-fusing framework, so that the key exposure can be overcome by fusing multiple keys. The key idea is to protect the randomness of keys through a family of functions, called key-fusing transformations (KFTs), each of which takes multiple independent keys as input, and can achieve at least the maximal randomness \cite{Dodis2011} over these inputted keys.  Even if part of the inputted keys is accidentally leaked to adversaries, the outputted key of KFT invariably maintains the maximal randomness, which enables the cryptosystem to be resistant to unpredicted or covert key exposure. The main contributions are summarized as follows:

\begin{itemize}
	\item A key-fusing framework to resist key exposure based on key-fusing transformations.
	\item A toy model for validating the correctness of the proposed framework by abstracting the PLKG mechanisms.
	\item Numerical results demonstrating the excellence of the proposed scheme in terms of secret outage probability (SOP).
\end{itemize}

The rest of this article is structured as follows. The motivation of this article is enlightened in the following section. 
We describe the proposed framework and verify its correctness and effectiveness through numerical results. We discuss the related work and open issues. The final section concludes the article.

\section{Motivation}
One of the basic problems in cryptography \cite{Maurer1993} is to prevent any adversary (referred to Eve) from obtaining useful information about the messages that are transmitted from a sender (Alice) to a receiver (Bob), even if Eve has a full access to the communication channel. According to Shannon's classic model, Eve can perfectly access the insecure channel, so the algorithms used in both encryption and decryption should be assumed to be public, i.e., the security of entire system is only dependent on the privacy or unpredictability of a cryptographic key shared between Alice and Bob.

From the adversary's point of view, a cryptographic key can be considered as a random variable in its key space. Once the key is leaked to the adversary, its randomness is lost, i.e., its entropy drops to zero. Since the Shannon entropy that describes the average unpredictability is not suitable for the case of formally evaluating cryptographic entropy, most contemporary scholars \cite{Dodis2011} \cite{Wu2019} adopt min-entropy, which is a conservative metric, to measure the difficulty or unpredictability of any adversary to guess a key. 
Unlike the Shannon entropy, min-entropy describes the worst-case unpredictability and is therefore required in modern cryptography.

In essence, the leakage of a key implies that its unpredictability is compromised by some adversary, so preventing key exposure is essentially equivalent to protecting the randomness of the key, i.e., ensuring its min-entropy large enough, which inspires the key idea of this article.

\section{The proposed key-fusing framework}\label{scheme}

\begin{figure}[htbp]
	\centering
	\includegraphics[width=0.45\textwidth]{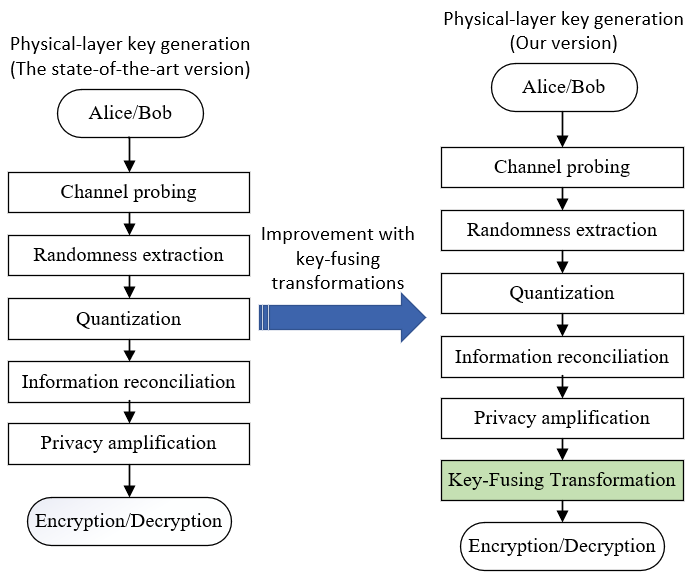}
	\caption{Improved version of physical-layer key generations with key-fusing transformations}
	\label{fig-frame}
\end{figure}

\subsection{Idea}

As shown in Fig.~\ref{fig-frame}, we take PLKG as an example to show how the proposed framework works.
Typically, PLKG enables two legitimate users, Alice and Bob, to extract a series of cryptography keys by performing five steps \cite{Jiao2019}, including channel probing, randomness extraction, quantization, reconciliation and privacy amplification.  And then the key can be further utilized for encryption and decryption with symmetric-key cryptography. It is generally believed that symmetric cryptography, such as AES, is able to resist quantum attacks, as long as the key size is large enough with a sufficiently high entropy. However, in theory, the secrecy of keys generated by PLKG cannot be guaranteed when eavesdroppers are within half a wavelength radius of legitimate users. In practice, an eavesdropper may covertly stalk legitimate users with passive or active attacks and the eavesdropper's location is usually unknown, which results in possible loss of information or even property due to key exposure.

The main idea of our framework is to increase the min-entropy as much as possible by fusing a set of keys. As shown in Fig.~\ref{fig-frame}, a special function, called key-fusing transformation (KFT), is performed on those keys generated by PLKG before they are utilized for encryption and decryption. Both Alice and Bob maintain a simple first-in first-out queue for managing this set of keys. The key feature of KFT is that it accepts a set of independent keys as input, and can output a new key that invariably achieves at least the maximal min-entropy over these inputted keys. That is to say, any eavesdropper cannot reduce the min-entropy of the KFT output to zero unless it has compromised all keys in the key set.  

\begin{figure}[htbp]
	\centering
	\includegraphics[width=0.45\textwidth]{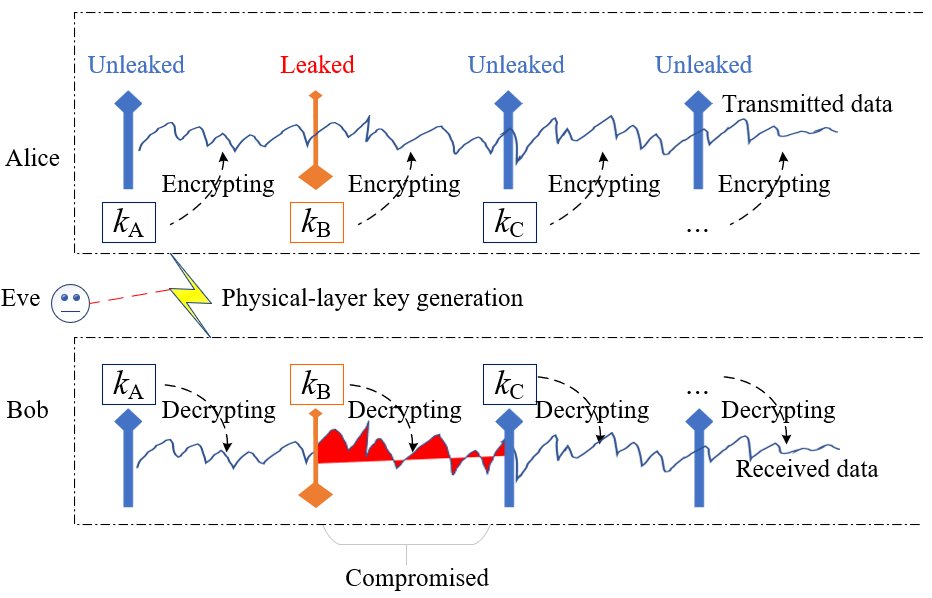}
	\caption{A typical scenario of eavesdropping with physical layer key generation}
	\label{fig-plkg}
\end{figure}

Fig.~\ref{fig-plkg} shows a typical scenario, in which Eve tries to covertly eavesdrop on the communication between Alice and Bob. $k_\text{A}$, $k_\text{B}$ and $k_\text{C}$ are three keys extracted by Alice and Bob through the PLKG mechanism and any one of them, say $k_\text{B}$, is leaked to Eve. By fusing these three keys with KFT, the exposure of  $k_\text{B}$ will be perfectly overcome with $k_\text{A}$ and $k_\text{C}$ that are in the same key set.

\subsection{Key-fusing transformations}


To our knowledge, the simplest key-fusing transformations are addition in modular arithmetic ($\text{MOD}$) and bitwise exclusive ($\oplus$) operator, both of which belong to the typical ARS (Addition/Rotation/$XOR$) primitives widely used in modern cryptography. Notice that applying one-to-one mapping on a random variable does not change its min-entropy, we can derive a multitude of KFT instances. The correctness and effectiveness of KFT can be validated through a toy model as illustrated in Section~\ref{validate}.

For a set of independent random variables, every KFT instance has the ability to maximize the min-entropy, though each of them  may lead to different results. Moreover, unless the min-entropy of an input variable is already close to the absolute maximum limited by the key size, the min-entropy of the KFT output is commonly much larger than that of every input, which implies that KFT actually fuses all inputted keys instead of simply choosing one of them, so that each input is perfectly hidden by the output---a beneficial property so called source privacy \cite{Aggarwal2014}. This property demands that the transformation does not expose any ``useful information'' about the source, so that the source can continuously work at full strength. Actually, because KFT allows each input to be in arbitrarily random distribution, it is impossible to guess part of the input variables from only the output.

\subsection{Fusion of multiple cryptographic keys}

We can extend KFT operations to a set of independent keys by iteratively performing KFT operations on each key in a KFT window. 
KFT enables the output to have the min-entropy greater than or equal to that of each input, which means that any potential leakage of a single key is hidden perfectly by the output in the sense of min-entropy, unless all keys are leaked concurrently. Correspondingly, any unleaked key contributes to the exposure resistance for all other keys in the KFT window.  Furthermore, the source privacy still holds because each input is permitted to be in arbitrarily random distribution and the range of its value cannot be narrowed down unless the KFT output and all other $ w-1 $ inputs are leaked together for a KFT window of size $w$.

In the case that the order of input variables is shuffled, whether the output remains identical is dependent on whether the adopted KFT instance follows the commutative and associative laws. Generally, all KFT instances meet the former, but not all satisfy the latter.

\section{Numerical Results}

In order to strictly verify the correctness  of the proposed scheme, we design a toy model to abstract the PLKG mechanisms.  The reason why we do not use a specific PLKG model is that any actual experiments and simulations cannot produce sufficiently rigorous proofs and verifications for modern cryptography.
\subsection{Toy model for PLKG}
The toy model is based on the following three assumptions.
\begin{enumerate}
	\item [S1:] Each key generated by PLKG is a random variable extracted from a ($n;l$)-key-source. 
	\item [S2:] Each key is independent of others.
	\item [S3:] The symmetric cryptography used for encryption and decryption is absolute secure if the key has a min-entropy greater than or equal to $l$, i.e., Eve can neither figure out  the key from the ciphertext, nor directly decrypt the ciphertext if  the key  is unleaked.
\end{enumerate}
Assumption S1 treats each key as a random variable. For a $n$-bit key extracted from ($n;l$)-key-source \cite{Aggarwal2014}, its min-entropy is less than $l$ if and only if it is leaked to Eve; Otherwise, its min-entropy must be greater than or equal to $l$.
Assumption S2 is reasonable because almost all research work related to PLKG requires testing the statistical independence of keys. Assumption S3 is reasonable because symmetric algorithms, such as AES, are generally considered strong enough to  resist even quantum computer's attacks as long as the key size is large enough.

\subsection{Validation of correctness} \label{validate}

Let us consider the scenario of eavesdropping with PLKG as shown in Fig. \ref{fig-plkg}. 
According to Assumption S1, we get that the min-entropy of $k_\text{A}$ and $k_\text{C}$ are both greater than or equal to  $l$, while that of  $k_\text{B}$ is less than $l$.
According to Assumption S3, communications encrypted with $k_\text{A}$ and $k_\text{C}$ are secure but the communication encrypted with $k_\text{B}$ is compromised. The main problem is that Alice and Bob do not even know that it is  $k_\text{B}$ that has been leaked, as the eavesdropping may be covert. 

To overcome this problem, our solution is to fuse $k_\text{A}$, $k_\text{B}$ and $k_\text{C}$ with KFT. Specifically, we get $k_\text{AB}$ from $k_\text{A}$ and $k_\text{B}$, and then $k_\text{ABC}$ from $k_\text{AB}$ and $k_\text{C}$. According to Assumption S3, what we need is to prove that the min-entropy of
$k_\text{ABC}$ is greater than or equal to  $l$.

\begin{figure}[htbp]
	\centering
	\includegraphics[width=0.45\textwidth]{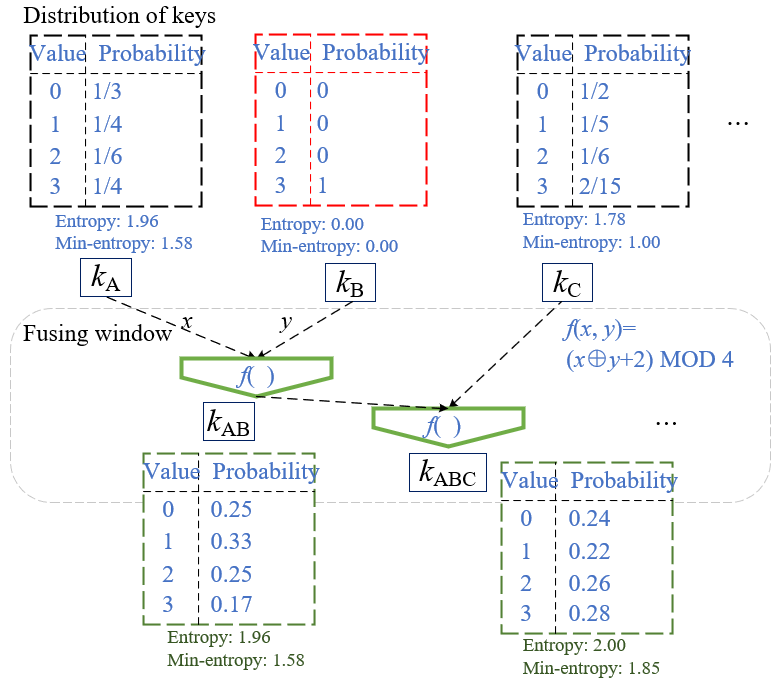}
	\caption{Numerical results for verifying the correctness of the proposed framework with 2-bit key model}
	\label{fig-kftproof}
\end{figure}

The magazine's style requirements do not allow us to provide complicated mathematical proofs, so we digitally verify the correctness of our model with a 2-bit key version, as shown in Fig.~\ref{fig-kftproof}, whose conclusion can be extended to any key size. 
In the 2-bit key model, the key size is 2, so each key can take one of four values 0, 1, 2, 3 with different probabilities. We set $k_\text{A}$ and $k_\text{C}$ to random variables in a distribution with probabilities \{1/3, 1/4, 1/6, 1/4\} and \{1/2, 1/5, 1/6, 2/15\}, respectively.   $k_\text{B}$ is leaked so its distribution is with probabilities \{0, 0, 0, 1\} for the assumed value of 3.

In Fig.~\ref{fig-kftproof}, the Shannon entropy and min-entropy of  $k_\text{B}$ are both zero because  $k_\text{B}$ is leaked. Thus, after applying the KFT instance $f( )$ to $k_\text{A}$ and $k_\text{B}$,  $k_\text{AB}$ still has the same min-entropy, 1.58, as $k_\text{A}$ but with different distribution. It should be noted that only two decimal digits of precision are kept for real numbers in the figure. It can be verified by computer program or manually that the min-entropy of $k_\text{ABC}$ is always greater than or equal to that of $k_\text{A}$, $k_\text{B}$ and $k_\text{C}$, i.e., the min-entropy of $k_\text{ABC}$ is always greater than or equal to $l$. Moreover, this conclusion always holds even if $k_\text{A}$ or $k_\text{C}$ takes another arbitrarily random distribution, or another KFT instance is used.


\begin{figure}[htbp]
	\centering
	\includegraphics[width=0.5\textwidth]{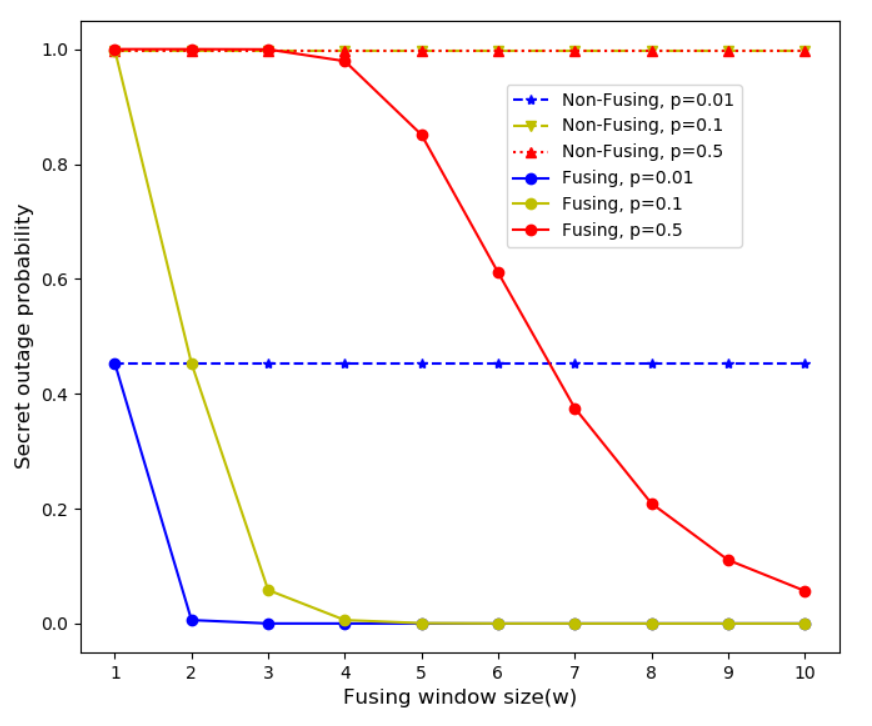}
	\caption{The effect of KFT on the secret outage probability}
	\label{fig-2B}
\end{figure}

\begin{figure}[htbp]
	\centering
	\includegraphics[width=0.5\textwidth]{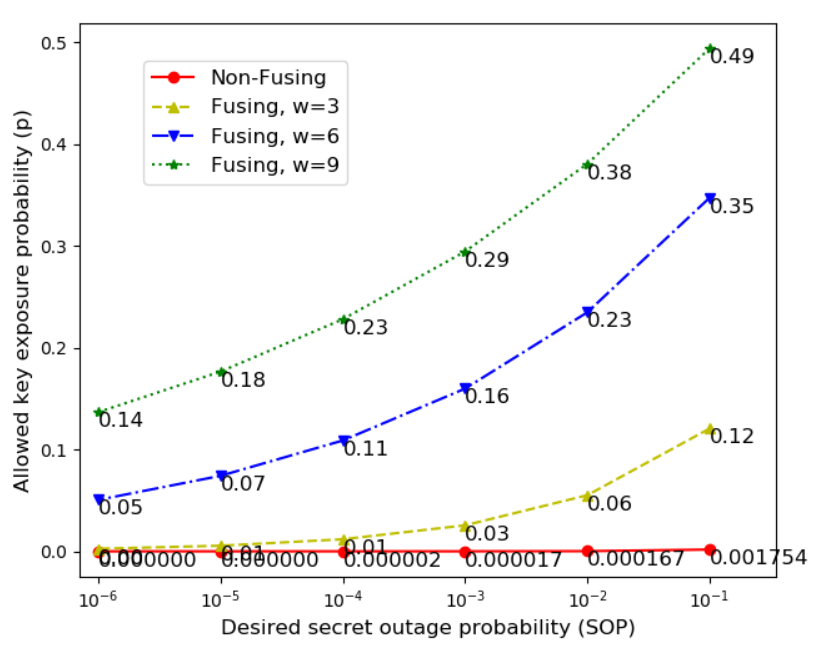}
	\caption{Allowed key exposure probability}
	\label{fig-exp}
\end{figure}

\subsection{Secret outage probability}
For a key $X$, its  exposure probability is defined as the probability that $X$ is leaked.
Secret outage probability (SOP) is the probability that the communication is compromised due to possible key exposure.  
Although it is impossible to eliminate the exposure probability of each key, KFT allows us to significantly reduce the SOP of the entire communication by fusing multiple keys.

Figure~\ref{fig-2B} demonstrates the effect of KFT on the SOP of the entire communication in which 60 keys are extracted. The approach to applying KFT is to displace every $w$ of those keys with the corresponding KFT output, where $w$ is the size of KFT window. In Fig.~\ref{fig-2B}, the two legends ``Non-Fusing'' and ``Fusing'' correspond to  
the SOP of the entire session without and with KFT activated, respectively. 
SOP are compared by varying the KFT window size  and the key exposure probability. It can be seen that the SOP with KFT activated 
is less than about $ 10^{-8} $ while 
the SOP without KFT is about 0.998 when the key exposure probability is 0.1  and the KFT window size is 10. Even for a rather high exposure probability, e.g., 0.5, the SOP with KFT activated 
becomes less than about 0.057 when the KFT window size is 10. 

Figure~\ref{fig-exp} demonstrates the relationship between allowed key exposure probability and the SOP to be expected. We see that, in order to meet the SOP of $10^{-6}$, the exposure probability of each key needs to be lower than $1.67\times 10^{-8}$, which is not practical for most implementations of PLKG. In contrast, by applying KFT with the KFT window size of 9, the same level of SOP can be achieved, as long as the key exposure probability is less than 0.14.

\section{The related work and open issues} \label{stateofart}

It is generally believed that most implementations of public-key cryptography, such as RSA, are vulnerable to quantum attacks, while lattice-based cryptography, quantum cryptography, physical-layer key generation are three of the main candidates for resisting attacks from quantum computers. 

\subsection{The threat of quantum computers}

At present, a multitude of lattice-based schemes, such as public key encryption and homomorphic encryption methods, have emerged. However, lattice-based cryptography is still vulnerable to key exposure and side channel attacks \cite{Chaudhary2019}. Moreover, to meet an acceptable security level, most implementations result in significant space or time complexity, which prevents the widespread adoption.

Quantum cryptography, also known as quantum key distribution (QKD), is on the basis of the quantum phenomena of non-locality and the violation of Bell inequalities. According to quantum physics, most eavesdropping methods will inevitably leave traces with quantum measurement characteristics, thus the action of eavesdropping will be noticed by Alice and Bob. Consequently, the eavesdropping can be prohibited by terminating the ongoing communication that is compromised. In this sense, quantum cryptography is absolutely secure and has resistance to quantum attacks. However, it still faces some grand challenges such as working distance, cost,  key-establishment rate, and practical security \cite{Arnon-Friedman2018}.

Physical-layer key generation is a promising wireless security technology, which exploits the intrinsic characteristics of the wireless channel to generate shared keys for legitimate users. Especially, 5G and beyond technologies offer many new opportunities for physical-layer key generation. For instance, massive MIMO-based beamforming adopted in 5G standard enables a high directionality that may be exploited to defend against co-located eavesdropping during key generation. However,  physical-layer key generation works securely in theory only when there are no eavesdroppers within half a wavelength radius of legitimate users \cite{Jiao2019}. Moreover, the concealment and unpredictability of eavesdropping make it difficult to avoid the key exposure problem completely.

\subsection{The threat of key exposure}

Key exposure is one of the main challenges faced by all cryptosystems that rely on the privacy of keys. Typically, key exposure problems are formulated as one-time attacks \cite{Naor2009a} or continuous leakage models \cite{Wu2019}.  The continuous leakage model often presumes that the leakage amount in each period is bounded, and the mainstream countermeasure is to update potentially leaked keys periodically by using fresh randomness \cite{Dodis2010, Zhou2019}. The system is assumed to be secure, as long as the key can be updated before the leakage exceeds a specific threshold.

However, due to the unpredictability and concealment of various attacks, this threshold may not be predetermined in many real-world applications. If the threshold is too large, the cryptosystem is in danger of being compromised. Otherwise, a great deal of precious entropy is wasted if the threshold is too small.

\subsection{Open issues}
In essence, our framework offers a way to protect the privacy of a single key by fusing multiple keys within a key-fusing window.  Whenever a sign of key exposure is detected, the security level of a cryptosystem can be boosted by enlarging the KFT window or introducing more fresh keys. As the KFT window size tends to infinity, the KFT output always goes up towards the absolute maximum min-entropy, $n$, for $n$-bit key space. In this sense, the proposed framework paves a promising way to achieving so-called information-theoretic security. An open question is whether or how fast the KFT output can reach the absolute maximum min-entropy, $n$.

Notice that almost all key generation mechanisms, such as public-key cryptography, quantum key distribution (QKD), and lattice-based cryptography, are supposed to satisfy Assumption S1 and S2, but these technologies have significantly different frameworks from PLKG. Therefore, it is another open problem how to integrate some of them in one system for better secret communication.

\section{Conclusion}
This article presents a key-fusing framework for resisting possible key exposure. Although there are a large number of key generation mechanisms, it is unreasonable that the effectiveness of almost all encryption algorithms is based on the privacy of a single key at every moment, which makes the cryptosystem extremely vulnerable to unpredictable or covert key exposure. To overcome this problem, we propose to employ a sliding KFT window for integrating various keys. Through a toy model abstracting the PLKG mechanisms, we verify the correctness of the proposed framework by numerical results and demonstrate that the secret outage probability is dramatically reduced.

Future research directions include in-depth exploration of the potential and limitation of KFT for various security missions \cite{Ni2019} in modern cryptosystems, such as encryption, authentication,  signature and constructing hash functions, where high randomness is strictly required. It is also an open problem how to integrate various key generation mechanisms in one system for achieving so-called information-theoretic security. These issues need further efforts invested. 

\section*{Acknowledgment}
This work was supported by the National Natural Science Foundation of China (61871076, 61273235, 61601097), the Fundamental Research Funds for the Central Universities of China (No. ZYGX2016J001), and the Defence Advance Research Foundation of China (No. 61400020109).

\ifCLASSOPTIONcaptionsoff
  \newpage
\fi



%
%
%

\bibliographystyle{IEEEtran}
\bibliography{IEEEabrv,security}

\end{document}